\documentclass{article}
\usepackage{amssymb,hyperref}
\usepackage[english]{babel}
\usepackage{amsmath, graphics}
\usepackage{geometry} 
\newcommand{\mathsym}[1]{{}}
\newcommand{\unicode}[1]{{}}

\def\noi{\noindent}
\def\nqq{\hspace{-2em}}

\def\barr{\left(\begin{array}}
\def\earr{\end{array}\right)}
\def\beq#1{\begin{equation}\label{#1}}
\def\eeq{\end{equation}}
\def\ber#1{\begin{eqnarray}\label{#1} &&\nqq}%   left alignment
\def\eer{\end{eqnarray}}

\newcommand{\bear}[1]{\begin{eqnarray}\label{#1}}
\newcommand{\ear}{\end{eqnarray}}

\catcode`\@=11 \@addtoreset{equation}{section}\catcode`\@=12
\newcommand{\N}{ {\mathbb N} }
\newcommand{\R}{ {\mathbb R} }

\newcommand{\fnm}{\footnotemark}
\newcommand{\fnt}{\footnotetext}

\begin{document}

 \vspace{15pt}

 \begin{center}
 \large\bf

 On flux integrals for generalized Melvin solution related 
  to simple finite-dimensional  Lie algebra

 \vspace{15pt}

 \normalsize\bf
       V. D. Ivashchuk\fnm[1]\fnt[1]{ivashchuk@mail.ru}$^{, a, b}$

 \vspace{7pt}

 \it (a) \ \ \ Center for Gravitation and Fundamental
 Metrology,  VNIIMS, 46 Ozyornaya Str., Moscow 119361, Russia  \\

 (b) \  Institute of Gravitation and Cosmology,
 Peoples' Friendship University of Russia (RUDN University),
 6 Miklukho-Maklaya Str.,  Moscow 117198, Russia \\

 \end{center}
 \vspace{15pt}

 \small\noi

 \begin{abstract}
 A generalized Melvin  solution for an arbitrary simple finite-dimensional Lie algebra $\cal G$  is considered. The  solution contains a metric, $n$ Abelian 2-forms and
 $n$ scalar fields, where $n$ is the rank of $\cal G$.
 It is governed by a set of $n$ moduli functions $H_s(z)$ obeying
 $n$ ordinary differential equations with certain boundary conditions
 imposed. It was conjectured earlier that these functions should
 be polynomials - the so-called fluxbrane polynomials. 
 These polynomials depend upon integration constants $q_s$, $s = 1,\dots,n$.
 In the case when the conjecture on the polynomial structure for the Lie algebra $\cal G$ is satisfied,
 it is proved that 2-form flux integrals $\Phi^s$ over a proper $2d$ submanifold are finite and obey  
 the relations:  $q_s \Phi^s =  4 \pi n_s  h_s$, where $h_s > 0$ are certain 
 constants (related to dilatonic coupling vectors) and $n_s$ are powers of  the polynomials,  which are components  of a twice dual  Weyl vector in the basis of simple (co-)roots,  $s = 1,\dots,n$. The main relations of the paper
  are valid for a solution corresponding to a finite-dimensional semi-simple Lie algebra $\cal G$.
 Examples of polynomials and fluxes for the Lie algebras  $A_1$, $A_2$, $A_3$,  $C_2$,  $G_2$ and  $A_1 +  A_1$  are presented.

 \end{abstract}

\large 

 \section{Introduction}

  In this paper we start with a  generalization of a  Melvin solution \cite{Melv},
   which was presented  earlier in ref. \cite{GI-09}. It appears in the  model 
  which contains a metric, $n$ Abelian 2-forms and  $l \geq n$ scalar fields.  
  This  solution is governed by a certain non-degenerate (quasi-Cartan) matrix 
  $(A_{s s'})$, $s, s' = 1, \dots, n$.  It is a special case of the so-called generalized fluxbrane  solutions from ref. \cite{Iflux}.  For fluxbrane solutions see  refs.
  \cite{GW}-\cite{IM-fb-14} and the references therein.  The appearance of fluxbrane solutions  was motivated  by superstring/$M$ theory.

  The generalized  fluxbrane solutions from ref. \cite{Iflux} are governed by 
  moduli functions   $H_s(z) > 0$ defined on the interval $(0, +\infty)$, where
   $z = \rho^2$ and $\rho$ is a radial variable. These functions 
  obey a set of  $n$ non-linear differential master equations governed by the matrix $(A_{s s'})$, equivalent to 
  Toda-like equations,  with  the following boundary conditions imposed:  $H_{s}(+ 0) = 1$,   $s = 1,\ldots,n$.  

 In this paper we assume that $(A_{s s'})$ is a Cartan matrix for some 
 simple finite-dimensional Lie algebra $\cal G$ of rank $n$ ($A_{ss} = 2$ for all $s$).
 According to a conjecture  suggested in ref. \cite{Iflux}, the
 solutions to the master equations with the  boundary conditions imposed 
 are  polynomials: 
  \beq{1.3}
  H_{s}(z) = 1 + \sum_{k = 1}^{n_s} P_s^{(k)} z^k,
  \eeq
  where $P_s^{(k)}$ are constants. Here
 $P_s^{(n_s)} \neq 0$  and 
 \beq{1.4}
 n_s = 2 \sum_{s' =1}^{n} A^{s s'},
 \eeq 
 where we denote $(A^{s s'}) = (A_{s s'})^{-1}$.
 The integers $n_s$ are components  of a twice dual
 Weyl vector in the basis of simple (co-)roots \cite{FS}.
 
The set of fluxbrane polynomials $H_s$ defines a 
special solution to  open Toda chain equations \cite{K,OP} corresponding 
to  a simple finite-dimensional Lie algebra $\cal G$ \cite{I-14}.
In refs. \cite{GI-09,GI} a program  (in Maple)  for the calculation of these polynomials for the classical series of Lie  algebras  ($A$-, $B$-, $C$- and $D$-series) was suggested.
It was pointed out in ref. \cite{Iflux} that the conjecture on polynomial structure of  $H_{s}(z)$ is valid for Lie  algebras of the $A$- and $C$-series.
In ref. \cite{BolIvas-17}    the conjecture from ref. \cite{Iflux} was verified for the Lie algebra $E_6$ and certain duality relations for six $E_6$-polynomials were proved. In Section 2 we present  the generalized  Melvin solution
from ref. \cite{GI-09}. In Section 3 we deal with the generalized  Melvin   solution for an arbitrary simple finite-dimensional Lie algebra $\cal G$.  
Here we calculate   2-form flux integrals $\Phi^s = \int_{M_{*}} F^s$, 
where  $F^s$ are 2-forms and $M_{*}$ is a certain $2d$ submanifold. 
These integrals (fluxes) are  finite when moduli functions are polynomials.
In Section 3 we consider examples of
fluxbrane polynomials and fluxes for the Lie algebras: 
$A_1$, $A_2$, $A_3$,  $C_2$,  $G_2$ and  $A_1 +  A_1$.

\section{The solutions}

We consider a  model governed by the action
 \beq{2.1}
 S=\int d^Dx \sqrt{|g|} \biggl \{R[g]-
 h_{\alpha\beta}g^{MN}\partial_M\varphi^{\alpha}\partial_N\varphi^{\beta}-\frac{1}{2}
 \sum_{s =1}^{n}\exp[2\lambda_s(\varphi)](F^s)^2 \biggr \}
 \eeq
 where $g=g_{MN}(x)dx^M\otimes dx^N$ is a metric,
 $\varphi=(\varphi^\alpha)\in\R^l$ is a set of scalar fields,
 $(h_{\alpha\beta})$ is a  constant symmetric non-degenerate
 $l\times l$ matrix $(l\in \N)$,    $ F^s =    dA^s
          =  \frac{1}{2} F^s_{M N}  dx^{M} \wedge  dx^{N}$
 is a $2$-form,  $\lambda_s$ is a 1-form on $\R^l$:
 $\lambda_s(\varphi)=\lambda_{s \alpha}\varphi^\alpha$,
 $s = 1,..., n$; $\alpha=1,\dots,l$.
 Here $(\lambda_{s \alpha})$, $s =1,\dots, n$, are dilatonic 
 coupling vectors.
 In (\ref{2.1})
 we denote $|g| =   |\det (g_{MN})|$, $(F^s)^2  =
 F^s_{M_1 M_{2}} F^s_{N_1 N_{2}}  g^{M_1 N_1} g^{M_{2} N_{2}}$, 
 $s = 1,\dots, n$.

 Here we start with a family of exact
solutions to field equations corresponding to the action
(\ref{2.1}) and depending on one variable $\rho$. The solutions
are defined on the manifold
 \beq{2.2}
  M = (0, + \infty)  \times M_1 \times M_2,
 \eeq
 where $M_1$ is a one-dimensional manifold (say $S^1$ or $\R$) and
 $M_2$ is a (D-2)-dimensional Ricci-flat manifold. The solution
 reads \cite{GI-09}
 \bear{2.30}
  g= \Bigl(\prod_{s = 1}^{n} H_s^{2 h_s /(D-2)} \Bigr)
  \biggl\{ w d\rho \otimes d \rho  +
  \Bigl(\prod_{s = 1}^{n} H_s^{-2 h_s} \Bigr) \rho^2 d\phi \otimes d\phi +
    g^2  \biggr\},
 \\  \label{2.31}
  \exp(\varphi^\alpha)=
  \prod_{s = 1}^{n} H_s^{h_s  \lambda_{s}^\alpha},
 \\  \label{2.32a}
  F^s=  q_s \left( \prod_{s' = 1}^{n}  H_{s'}^{- A_{s
  s'}} \right) \rho d\rho \wedge d \phi,
  \ear
 $s = 1,\dots, n$; $\alpha = 1,\dots, l$, where $w = \pm 1$, $g^1 = d\phi \otimes d\phi$ is a
  metric on $M_1$ and $g^2$ is a  Ricci-flat metric on
 $M_{2}$.  Here $q_s \neq 0$ are integration constants,  
 $q_s = - Q_s$ in the notations of ref. \cite{GI-09}, $s = 1,\dots, n$.

 The functions $H_s(z) > 0$, $z = \rho^2$, obey the master equations
\beq{1.1}
  \frac{d}{dz} \left( \frac{ z}{H_s} \frac{d}{dz} H_s \right) =
   P_s \prod_{s' = 1}^{n}  H_{s'}^{- A_{s s'}},
  \eeq
 with  the following boundary conditions
 \beq{1.2}
   H_{s}(+ 0) = 1,
 \eeq
 where
 \beq{2.21}
  P_s =  \frac{1}{4} K_s q_s^2,
 \eeq
 $s = 1,\dots,n$.  The boundary condition (\ref{1.2}) guarantees the absence 
 of a conic singularity (in the metric  (\ref{2.30})) for $\rho =  +0$.
 
 The parameters  $h_s$  satisfy the relations
  \beq{2.16}
  h_s = K_s^{-1}, \qquad  K_s = B_{s s} > 0,
  \eeq
 where
 \beq{2.17}
  B_{ss'} \equiv
  1 +\frac{1}{2-D}+  \lambda_{s \alpha} \lambda_{s' \beta}   h^{\alpha\beta},
  \eeq
 $s, s' = 1,\ldots, n$, with $(h^{\alpha\beta})=(h_{\alpha\beta})^{-1}$.
 In relations above we denote 
 $\lambda_{s}^{\alpha} = h^{\alpha\beta}  \lambda_{s \beta}$
 and
 \beq{2.18}
  (A_{ss'}) = \left( 2 B_{s s'}/B_{s' s'} \right).
 \eeq
 The latter is the so-called quasi-Cartan matrix. 
 
 We note that the constants   $B_{s s'}$ and $K_s = B_{s s}$
 have a certain mathematical sense.  They are related to scalar products of certain vectors $U^s$ (brane 
 vectors, or $U$-vectors),  which belong  to a certain linear space (``truncated target space'', for our problem it has dimension $l+2$), i.e.  $B_{s s'} =(U^s,U^{s'})$ and  $K_s = (U^s,U^{s})$ \cite{IMJ,IM-top,IMsigma}. The scalar  products of such a type are  of physical significance, since they appear for various solutions with branes,
 e.g. black branes, $S$-branes, fluxbranes etc. Several physical parameters in multidimensional models with branes, e.g. the Hawking-like temperatures and the entropies of black holes and branes, PPN parameters, Hubble-like parameters,
  fluxes etc., contain such scalar products; see \cite{IM-top,IMsigma} and Section 3 of this paper. The relation (\ref{2.18}) defines generalized intersection rules for branes which were suggested  in \cite{IMJ}.   The constants $K_s$ are invariants of dimensional reduction.  It is wel known, see \cite{IMsigma} and the references therein, that $K_s = 2$ for branes  in numerous supergravity models, e.g. in dimensions $D = 10,11$.
   
 It may be shown that if the matrix $(h_{\alpha\beta})$
 has an Euclidean signature and  $l \geq n$, 
 and  $(A_{ss'})$ is a Cartan matrix 
 for a simple Lie algebra $\cal G$ of rank $n$,
 there exists a set of co-vectors 
 $\lambda_1, \dots, \lambda_n$  obeying (\ref{2.18}) (for $l = n$ 
 see Remark 1 in the next section.).
 Thus the solution is valid at least when $l \geq n$
 and the matrix $(h_{\alpha\beta})$ is  positive-definite.

The solution under consideration is a special case of the
fluxbrane (for $w = +1$,  $M_1 = S^1$) and $S$-brane
($w = -1$) solutions from \cite{Iflux} and \cite{GIM}, respectively.

If $w = +1$ and the (Ricci-flat) metric $g^2$ has a
pseudo-Euclidean signature, we get a multidimensional 
generalization of   Melvin's solution \cite{Melv}. 
 
In our notations  Melvin's solution (without scalar field) corresponds to $D = 4$,
$n = 1$, $l =0$, $M_1 = S^1$ ($0 < \phi <  2 \pi$),  $M_2 = \R^2$,
$g^2 = -  dt \otimes dt + d x \otimes d x$
and ${\cal G} = A_1$. 
  
For $w = -1$ and $g^2$ of Euclidean signature we obtain a cosmological solution
with a horizon (as $\rho = + 0$) if $M_1 = \R$ ($ - \infty < \phi < + \infty$).

\section{Flux integrals  for  a simple finite-dimensional  Lie algebra}

Here we deal with the solution which corresponds to a simple finite-dimensional Lie algebra ${\cal G}$, i.e. 
the matrix   $A =  (A_{ss'})$ is coinciding with the Cartan matrix of this Lie algebra.
We put also  $n = l$, $w = + 1$ and $M_1 = S^1$,  $h_{\alpha\beta} = \delta_{\alpha \beta}$ 
and denote  $(\lambda_{s a }) =  (\lambda_{s}^{a}) = \vec{\lambda}_{s}$, $s = 1, \dots, n$.

  Due to (\ref{2.16})-(\ref{2.18}) we get
    \begin{equation}
           \label{3.17}
    K_s =   \frac{D - 3}{D -2} +  \vec{\lambda}_{s}^2,
      \end{equation} 
    $h_s =  K_s^{-1}$,  and  
      \beq{3.18}
        \vec{\lambda}_{s} \vec{\lambda}_{l} = 
            \frac{1}{2} K_l A_{sl}  - \frac{D - 3}{D -2} \equiv \Gamma_{sl},
      \eeq    
     $s,l = 1, \dots, n$. ((\ref{3.17}) is a special case of  (\ref{3.18}). )
         
          It follows from  (\ref{2.16})-(\ref{2.18}) that 
         \beq{3.26}
          \frac{h_i}{h_j} = \frac{K_{j}}{K_{i}} = \frac{B_{jj}}{B_{ii}} = 
          \frac{B_{ji}}{B_{ii}} \frac{B_{jj}}{B_{ij}} = \frac{A_{ji}}{A_{ij}}
           \eeq
         for any $i \neq j$ obeying $A_{ij}, A_{ji} \neq 0$; $i,j = 1, \dots,n$.
      It may be readily shown from (\ref{3.26}) that  the ratios 
      $\frac{h_i}{h_j} = \frac{K_{j}}{K_{i}}$ are fixed numbers for any given Cartan 
      matrix $({A_{ij}})$ of a simple (finite-dimensional) Lie algebra ${\cal G}$.
      (This follows from (\ref{3.26}) and the  connectedness of the Dynkin diagram of a simple Lie algebra.) 
      The ratios (\ref{3.26}) may be written as follows:
       \beq{3.27}
        \frac{h_i}{h_j} = \frac{K_{j}}{K_{i}} = \frac{r_j }{r_i} 
         \eeq
        $i \neq j$, where $r_i = (\alpha_{i}, \alpha_{i})$ is the length squared  of a simple root $\alpha_{i}$ corresponding to the Lie algebra ${\cal G}$. Here  we use the notations
        $A_{ij} = 2 (\alpha_{i}, \alpha_{j})/(\alpha_{j}, \alpha_{j})$;
        $i,j = 1, \dots,n$. Relation (\ref{3.27}) implies 
      \beq{3.27a}
       K_{i} = \frac12 K r_i,  
      \eeq
     $i = 1, \dots,n$, where $K > 0$.
     (For simply laced ($A,D,E$) Lie algebras all $r_i$ are equal.) 
     
          {\bf Remark 1.} 
     {\em For large enough $K$ in (\ref{3.27a}) there exist vectors 
     $\vec{\lambda}_s$  obeying   (\ref{3.18}) (and hence (\ref{3.17})).
     Indeed, the matrix $(\Gamma_{sl})$ is positive-definite
     if $K > K_{*}$,  where $K_{*}$ is some positive number. Hence there exists
     a matrix $\Lambda$, such that $\Lambda^{T}\Lambda = \Gamma$. We put
      $(\Lambda_{as}) = (\lambda_{s}^a)$ and get the set of vectors obeying
      (\ref{3.18}).}

      Now let us consider the oriented $2$-dimensional manifold 
      $M_{*} =(0, + \infty)  \times S^1$. The flux integrals
       \beq{3.19}
       \Phi^s = \int_{M_{*}} F^s =  
        \int_{0}^{+ \infty} d \rho \int_{0}^{2 \pi} d \phi \ \rho {\cal B}^s(\rho^2)
        = 2 \pi \int_{0}^{+ \infty} d \rho \ \rho {\cal B}^s(\rho^2) ,
       \eeq    
       where 
        \begin{equation}
           \label{3.16}
           {\cal B}^s(\rho^2) =   q_s  \prod_{l = 1}^{n}  (H_{l}(\rho^2))^{- A_{s l}},
        \end{equation} 
       are convergent for all $s$, if the conjecture for the Lie algebra ${\cal G}$ 
       (on polynomial structure  of moduli functions $H_s$)   is obeyed 
       for the  Lie algebra ${\cal G}$ under consideration.
       
        Indeed, due to the polynomial assumption  (\ref{1.3}) we have 
      \beq{3.20}
       H_s(\rho^2) \sim C_s \rho^{2n_s},   \qquad C_s = P_s^{(n_s)},
            \eeq 
      as $\rho \to + \infty$; $s =1, \dots, n$. From (\ref{3.16}), (\ref{3.20}) and 
      the equality $\sum_{1}^{n} A_{s l} n_l = 2$,
    following from (\ref{1.4}), we get
      \beq{3.21}
           {\cal B}^s(\rho^2) \sim  q_s C^s \rho^{-4},  \quad  C^s =  \prod_{l = 1}^{n} C_l^{-A_{sl}},              
       \eeq 
       and hence the integral (\ref{3.19}) is convergent for any $s =1, \dots, n$.

        By using  the master equations (\ref{1.1})  we obtain
      \bear{3.24}
      \int_{0}^{+ \infty} d \rho \rho {\cal B}^s(\rho^2) =  q_s P_s^{-1}
      \frac12 \int_{0}^{+ \infty} d z  
      \frac{d}{dz} \left( \frac{ z}{H_s} \frac{d}{dz} H_s \right)
      \\ \nonumber
      =     \frac12 q_s P_s^{-1} \lim_{z \to + \infty}   
              \left( \frac{ z}{H_s} \frac{d}{dz} H_s \right) =
               \frac12 n_s q_s P_s^{-1},          
      \ear
   which implies (see (\ref{2.21}))
            \beq{3.25}
         \Phi^s =   4 \pi n_s q_s^{-1} h_s,  
         \eeq
   $s =1, \dots, n $. 
  
  Thus, any flux $\Phi^s$ depends upon one integration constant  $q_s \neq 0$, 
  while the integrand form $F^s$ depends upon all constants: $q_1, \dots, q_n$.
   
  We note that  for $D =4$ and 
   $g^2 = -  dt \otimes dt + d x \otimes d x$, $q_s$ is coinciding  with 
  the value of the $x$-component of the $s$th magnetic field on the axis of symmetry. 
  
  In the case of the Gibbons-Maeda dilatonic generalization of the Melvin solution,   corresponding to  
  $D = 4$, $n = l= 1$   and ${\cal G} = A_1$ \cite{GM}, the flux from (\ref{3.25}) ($s=1$)  is in agreement  
  with that considered in ref. \cite{KKR}. For the Melvin's case and some higher dimensional 
  extensions (with ${\cal G} = A_1$)    see also ref. \cite{GutSt}.
   
 Due to (\ref{3.27})  the ratios 
    \beq{3.28}
    \frac{q_i \Phi^i}{q_j \Phi^j} =  \frac{n_i h_i}{n_j h_j} =  \frac{n_i r_{j} }{n_j r_{i}}  
    \eeq
  are fixed numbers  depending upon the  Cartan  matrix $({A_{ij}})$ of a simple finite-dimensional Lie algebra 
  ${\cal G}$.
 
 {\bf Remark 2.}
 {\em The  relation for flux integrals  (\ref{3.25}) is also valid when the matrix  
 $(A_{ss'})$   is a Cartan matrix of a finite-dimensional semi-simple Lie algebra ${\cal G} = {\cal G}_1 \oplus \cdots \oplus {\cal G}_k$, where ${\cal G}_1,  \dots, {\cal G}_k$ are simple Lie (sub)algebras. In this case the Cartan matrix $({A_{ij}})$ has a block-diagonal
 form, i.e. $({A_{ij}}) = {\rm diag} (({A^{(1)}_{i_1 j_1}}), \cdots, ({A^{(k)}_{i_k j_k}}))$,
 where $({A^{(a)}_{i_a j_a}})$ is the Cartan matrix of the Lie algebra ${\cal G}_a$,
 $a = 1, \dots, k$. The set of polynomials in this case splits in the direct union
 of sets of polynomials corresponding to the Lie algebras  ${\cal G}_1,  \dots, {\cal G}_k$.
 Relations (\ref{3.27}) and (\ref{3.28}) are valid, when the indices $i, j$ correspond to one $a$th block, $a = 1, \dots, k$. The quantities $q_i \Phi^i$ and $q_j \Phi^j$ corresponding  to different blocks are independent.
 Relation  (\ref{3.27a}) should be replaced by
  \beq{3.29}
         K_{i_a} = \frac12 K^{(a)} r_{i_a},  \quad K^{(a)} > 0,
        \eeq
       for any index $i_a$  corresponding to $a$-th block;   $a = 1, \dots, k$.  
       The existence of dilatonic coupling vectors $\vec{\lambda}_s$  obeying   (\ref{3.18}) (and (\ref{3.17}))
       just follows from the arguments of  Remark $1$, if we put all $K^{(a)} = K > 0$.} 
   
  The manifold $M_{*} =(0, + \infty)  \times S^1$ is isomorphic to
   the manifold $\R^2_{*} = \R^2 \setminus \{ 0 \}$.  The solution (\ref{2.30})-(\ref{2.32a}) may be understood (or rewritten by pull-backs) as defined on the manifold $\R^2_{*} \times M_2$,  where coordinates $\rho$, $\phi$ are 
   understood as coordinates on  $\R^2_{*}$. They  are not globally defined. One should consider two charts   with  coordinates $\rho$, $\phi= \phi_1$ and $\rho$, $\phi= \phi_2$, where $\rho > 0$, $0 < \phi_1 < 2 \pi$ and  $- \pi < \phi_2 <  \pi$. Here $\exp(i \phi_1 ) = \exp(i \phi_2)$. In both cases we have  $x = \rho \cos \phi$ and
    $y = \rho \sin \phi $, where $x, y$ are standard coordinates of $\R^2$.
    Using the identity $ \rho  d\rho \wedge  d\phi = d x \wedge  dy$ we get
    \beq{3.30}
     F^s=  q_s  \prod_{s' = 1}^{n}  
      (H_{s'}(x^2 + y^2))^{- A_{s s'}}   dx \wedge d y,
    \eeq
    $s =1, \dots, n $.   The 2-forms (\ref{3.30}) are well defined on $\R^2$. Indeed, 
    due to conjecture from ref. \cite{Iflux} any  polynomial $H_{s}(z)$ is a smooth function
    on $\R = (- \infty, + \infty)$ which obeys $H_{s}(z) > 0$ for $z \in (- \varepsilon_s, + \infty)$, where 
    $\varepsilon_s > 0$. This is valid since due to conjecture from ref. \cite{Iflux} $H_{s}(z) > 0$ for $z > 0$
    and $H_{s}(+0) = 1$. Thus, $\left( \prod_{s' = 1}^{n} (H_{s'}(x^2 + y^2))^{- A_{s s'}} \right)$
    is a smooth function since it is a composition of two well-defined smooth functions  
    $\left( \prod_{s' = 1}^{n} (H_{s'}(z))^{- A_{s s'}} \right)$ and $z = x^2 + y^2$. 
           
       Now we show that there exist 1-forms $A^s$ obeying $F^s = dA^s$ which are globally defined on $\R^2$.
    We start with the open submanifold  $\R^2_{*}$. The  1-forms 
    \beq{3.31}
     A^s = \left( \int_{0}^{\rho}  d \bar{\rho} \bar{\rho} {\cal B}^s(\bar{\rho}^2)  \right) d \phi
         = \frac{1}{2}  \left( \int_{0}^{\rho^2} d \bar{z}  {\cal B}^s(\bar{z})  \right) d \phi
    \eeq
    are well defined on $\R^2_{*}$ (here $d\phi = (x^2 + y^2)^{-1} ( - y dx + x dy)$) and obey $F^s = dA^s$,
    $s = 1, \dots, n $. Using the master equation (\ref{1.1}) we obtain
     \bear{3.32}
         A^s = \frac{q_s}{2 P_s}  \left( \int_{0}^{\rho^2} d \bar{z} 
    \frac{d}{d \bar{z}}  \left( \frac{ \bar{z}}{H_s(\bar{z})} \frac{d}{d\bar{z}} H_s ( \bar{z}) 
    \right) \right) d \phi
         \\ \nonumber
     =  \frac{2 h_s}{q_s} \frac{H^{'}_{s} (\rho^2)}{H_s(\rho^2)} \rho^2 d \phi,  
        \ear
    $s = 1, \dots, n $. Here $H{'}_s = \frac{d}{dz} H_s$. Due to relation $\rho^2 d \phi = - y dx + x dy $,
    we obtain 
     \beq{3.33}
             A^s   =  \frac{2 h_s}{q_s} \frac{H^{'}_{s} (x^2 + y^2)}{H_s(x^2 + y^2)} (- y dx + x dy),  
     \eeq
      $s = 1, \dots, n $. The 1-forms (\ref{3.33}) are well defined smooth  1-forms on $\R^2$.
      
      We note that in the case of the Gibbons-Maeda solution \cite{GM} corresponding to  $D = 4$, $n = l= 1$ 
      and ${\cal G} = A_1$ the gauge potential from (\ref{3.32}) coincides (up to notations) 
      with that considered in ref. \cite{DGGidH}.
                         
     Now we verify our result (\ref{3.25}) for flux integrals by using the relations for the 1-forms $A^s$. 
     Let us consider a $2d$ oriented manifold (disk) $D_R = \{ (x,y): x^2 + y^2 \leq R^2 \}$ with the boundary
     $\partial D_R = C_R = \{ (x,y): x^2 + y^2 = R^2 \}$. $C_R$ is a circle of radius $R$. It is
     an $1d$ oriented manifold  with the orientation (inherited from that of $D_R$) obeying the relation $\int_{C_R} d\phi = 2 \pi$. 
     Using the Stokes-Cartan theorem we get
     \beq{3.34}
         \int_{D_{R}} F^s=    \int_{D_{R}} d A^s=   \int_{C_{R}}  A^s
          =  \frac{4 \pi h_s}{q_s} \frac{H^{'}_{s} (R^2)}{H_s(R^2)} R^2,  
     \eeq
     $s = 1, \dots, n $. 
     By using the asymptotic relation (\ref{3.20}) we find
     \beq{3.35}
        \lim_{R \to + \infty}  \int_{D_{R}} F^s  =  \frac{4 \pi h_s n_s}{q_s},  
          \eeq
          $s = 1, \dots, n $, in agreement with (\ref{3.25}).   
     
     {\bf Remark 3.} {\em  We note (for a completeness) that the metric and scalar fields
     for our solution with $w = +1$ and $l = n$ 
     can be extended  to the manifold $\R^2 \times M_2$. 
     Indeed, in the coordinates $x, y$ the metric (\ref{2.30}) and scalar fields
     (\ref{2.31}) read as follows          
      \bear{3.36}
        g= \Bigl(\prod_{s = 1}^{n} H_s^{2 h_s /(D-2)} \Bigr)
        \biggl\{  d x \otimes d x  + d y \otimes d y + 
         f  (- y dx + x dy)^2 +   g^2  \biggr\},
       \\  \label{3.37}
        \varphi^{a}=
        \sum_{s = 1}^{n} h_s  \lambda_{s}^{a} \ln H_s,
          \ear
     $a =1, \dots, l $. Here $H_s = H_s(x^2 + y^2)$, $s = 1, \dots, n $, and $f = f(x^2 + y^2)$,
     where
     \beq{3.38}
     f(z) = \left(\Bigl(\prod_{s = 1}^{n} (H_s(z))^{-2 h_s} \Bigr) - 1 \right) z^{-1},
     \eeq
     for $z \neq 0$ and $f(0)= \lim_{z \to 0} f(z)$ (the limit does exist).  The function $f(z)$ is smooth in 
     in the interval  $ (- \varepsilon, + \infty)$ for some $\varepsilon > 0$. Indeed, it is
     smooth  in  the interval $ (0, + \infty)$ and  holomorphic in the domain
     $\{z | 0 < |z| < \varepsilon  \}$ for a small enough  $\varepsilon > 0$. Since the limit
     $\lim_{z \to 0} f(z)$ does exist the function $f(z)$ is  holomorphic in the disc
          $\{z | |z| < \varepsilon  \}$ and hence it is smooth in the interval $ (- \varepsilon, + \infty)$.
     This implies that the metric is  smooth on the manifold $\R^2 \times M_2$. 
     (See the text after the formula (\ref{3.30}).)
     The scalar fields are also smooth on $\R^2 \times M_2$.   }  
     
  \section{Examples }
 
  Here we present  fluxbrane polynomials corresponding to the Lie algebras
 $A_1$, $A_2$, $A_3$,  $C_2$,  $G_2$, $A_1 +  A_1$  and related fluxes.
  Here as in \cite{I-14}  we use other parameters $p_s$   instead of $P_s$:
    \beq{4.g}
     p_s =  P_s/n_s,
    \eeq
    $s = 1, \ldots, n$. 
    %This is done for avoiding big denominators in $P_s^{(k)}$.

   {\bf $A_1$-case.} The simplest example occurs in the case of
  the Lie algebra $A_1 = sl(2)$. Here $n_1 = 1$.
   We get  \cite{Iflux}
  \beq{A.1}
    H_{1}  = 1 + p_1 z
   \eeq
    and 
   \beq{A1f}
   \Phi^1 =   4 \pi  q_1^{-1} h_1,
   \eeq
   which is also valid 
   for Melvin's solution with $D = 4$ and $h_1 = 2$.  
  
  {\bf $A_2$-case.} For the Lie algebra $A_2 = sl(3)$ with the
 Cartan matrix 
 
  \beq{A.5}
     \left(A_{ss'}\right)=
   \left( \begin{array}{*{6}{c}}
      2 & -1\\
      -1& 2\\
     \end{array}
  \right)\quad \eeq
  we have \cite{Iflux,GIM,I-14}  $n_1 = n_2 =2$ and
  \bear{A.6}
    H_{1} = 1 + 2 p_1 z +  p_1 p_2 z^{2}, \\
   \label{A.7}
    H_{2} = 1 + 2 p_2 z +  p_1 p_2 z^{2}.
   \ear
    We get in this case
   \beq{A2f}
      (\Phi^1, \Phi^2) =   8 \pi h  (q_1^{-1},q_2^{-1}),
   \eeq
   where $h_1 = h_2 = h$.
 
  {\bf $A_3$-case.} The polynomials for the $A_3$-case read as
  follows \cite{GI,I-14}
 
  \bear{A.8}
    H_{1} = 1 +  3 p_1 z + 3 p_1 p_2 z^{2} +  p_1 p_2 p_3 z^{3}, \\
  \label{A.9}
    H_{2} = 1 + 4 p_2 z + 3 \Bigl(  p_1 p_2 +
       p_2 p_3 \Bigr) z^{2} + 4 p_1 p_2 p_3 z^{3}
     +  p_1 p_2^{2} p_3 z^{4},\\
  \label{A.10}
   H_{3} = 1 + 3 p_3 z + 3 p_2 p_3 z^{2} +  p_1 p_2 p_3 z^{3}.
  \ear
   Here we have  $(n_1, n_2, n_3) = (3,4,3)$ 
   and 
   \beq{A3f}
   (\Phi^1, \Phi^2,\Phi^3) =   4 \pi h  (3q_1^{-1},4q_2^{-1}, 3q_3^{-1})
   \eeq
   with $h_1 = h_2 = h_3 = h$.

 {\bf $C_2$-case.}
  For the Lie algebra $C_2 = so(5)$ with the Cartan matrix 
 
  \beq{C.1}
     \left(A_{ss'}\right)=
   \left( \begin{array}{*{6}{c}}
      2 & -1\\
      -2& 2\\
  \end{array}
  \right)\quad
  \eeq
  we get   $n_1 = 3$ and $n_2 = 4$.
   For $C_2$-polynomials we obtain \cite{GIM,I-14}
  \bear{C.2}
    H_1 = 1+ 3 p_1 z+ 3 p_1 p_2 z^2 + p_1^2 p_2 z^3,
                \\ \label{C.3}
    H_2 = 1+ 4 p_2 z+ 6 p_1 p_2 z^2 + 4 p_1^2 p_2 z^3
         +  p_1^2 p_2^2 z^4.
  \ear
  In this case we find 
 \beq{Cf}
       (\Phi^1, \Phi^2) =   4 \pi  (3 h_1 q_1^{-1}, 4 h_2 q_2^{-1})
    \eeq
    where $h_1 = 2 h_2$.
   
  { \bf $G_2$-case. }
  For the Lie algebra $G_2$ with the Cartan matrix
 
 \beq{G.1}
     \left(A_{ss'}\right)=
   \left( \begin{array}{*{6}{c}}
      2 & -1\\
      -3& 2\\
  \end{array}
  \right)\quad
  \eeq
  we get   $n_1 = 6$ and $n_2 = 10$.
 In this case  the fluxbrane polynomials read \cite{GIM,I-14}
  \vspace{15pt}
 \bear{G.2}
   H_{1} = 1+ 6 p_1 z+ 15 p_1 p_2 z^2 + 20 p_1^2 p_2 z^3 + \\ \nonumber
     15 p_1^3 p_2 z^4 + 6 p_1^3 p_2^2 z^5 + p_1^4 p_2^2 z^6 ,
 \\ \label{G.3}
    H_2 =  1+  10 p_2 z + 45 p_1 p_2 z^2  +  120 p_1^2 p_2 z^3
   +  p_1^2 p_2( 135 p_1 + 75 p_2) z^4 \\ \nonumber
                 + 252 p_1^3 p_2^2 z^5
    + p_1^3 p_2^2 \biggl(75 p_1 + 135 p_2 \biggr)z^6   +  120 p_1^4 p_2^3  z^7
    \\ \nonumber
    + 45 p_1^5 p_2^3 z^8 +  10 p_1^6 p_2^3 z^9
       + p_1^{6} p_2^{4}  z^{10}.
   \ear
  \vspace{15pt}
  We are led to relations
  \beq{Gf}
         (\Phi^1, \Phi^2) =   4 \pi  (6 h_1 q_1^{-1}, 10 h_2 q_2^{-1})
      \eeq
      where $h_1 =  3 h_2$.
  
  {\bf ($A_1 + A_1$)-case.} For  semi-simple Lie algebra  $A_1 + A_1$ we obtain  
   $n_1 = n_2 = 1$, 
       \beq{AA.1}
      H_{1} = 1 + p_1 z, \quad  H_{2} = 1 + p_2 z, 
     \eeq
     and 
     \beq{AA1f}
     (\Phi^1, \Phi^2)  =   4 \pi  (q_1^{-1} h_1, q_2^{-1} h_2),
     \eeq
     where $h_1$ and $h_2$ are independent, as well as the quantities
     $q_1 \Phi^1$ and $q_2 \Phi^2$.

  \section{\bf Conclusions}
  
    Here we have considered a multidimensional generalization of Melvin's
    solution corresponding to a simple finite-dimensional Lie algebra ${\cal G}$. 
    We have assumed that the solution is governed by a set of $n$  fluxbrane polynomials $H_s(z)$, $s =1,\dots,n$. 
    These polynomials  define special solutions to open Toda chain equations corresponding 
    to the Lie algebra  ${\cal G}$.
     
    The polynomials $H_s(z)$ depend also upon parameters $q_s$, which   
    are coinciding for $D =4$  (up to a sign) with the values  of  colored 
    magnetic fields on the axis of symmetry. 
      
    We have  calculated $2d$ flux integrals $\Phi^s = \int F^s$ , $s =1, \dots, n$.
    Any flux  $\Phi^s$ depends only upon one parameter $q_s$, while the integrand  $F^s$ depends upon all parameters $q_1, \dots, q_n$.
    The  relation for flux integrals  (\ref{3.25}) is also valid when the matrix  
    $(A_{ss'})$   is a Cartan matrix of a finite-dimensional semi-simple Lie algebra $\cal G$. 
     
    Here we have considered   examples of polynomials and fluxes for the Lie algebras  $A_1$, $A_2$, $A_3$,  $C_2$,  $G_2$ and  $A_1 +  A_1$.
    The approach of this paper will be used for a calculation of certain flux integrals
    for forms $F^s$  of arbitrary ranks corresponding to certain  fluxbrane solutions     
    (of electric type by $p$-brane notation or magnetic type by fluxbrane classification
     \fnm[2]\fnt[2]{We remind the reader that an electric (magnetic) $p$-brane corresponds to a magnetic (electric)
         $F(D-3 -p)$ fluxbrane, see ref. \cite{Iflux} and the references therein.})                             
     governed by fluxbrane polynomials  \cite{Ivas-flux-17}.     
   
    An open problem is to  find the fluxes for the solutions which are related  to  infinite-dimensional Lorentzian Kac-Moody algebras, e.g. hyperbolic ones \cite{Kac,HPS}. In this case one should deal with 
    phantom scalar fields in the model (\ref{2.1}) and non-polynomial solutions to eqs. (\ref{1.1}). Another possibility is to study the convergence of flux integrals for 
    non-polynomial solutions for moduli functions corresponding to  non-Cartan matrices  $(A_{ss'})$ 
    (e.g. for the model with two $2$-forms from ref. \cite{ABI}).
            
 \newpage 
 
    \begin{center}
    {\bf Acknowledgments}
    \end{center}
  
   This work was supported in part by the Russian Foundation for
   Basic Research Grant No. 16-02-00602 and by the Ministry of Education of the Russian Federation (the  Agreement Number 02.a03.21.0008 of 24 June 2016).

 \end{document}